\documentclass[aps,pra,twocolumn,amsmath,amssymb,preprintnumbers,showpacs]{revtex4}
\usepackage{graphicx}
\usepackage{dcolumn}

\begin{document}


\title{Non-Markovian weak coupling limit of quantum Brownian motion}

\author{Sabrina Maniscalco}
\affiliation{Department of Physics and Astronomy, University of Turku, FI-20014 Turku, Finland}
\author{Jyrki Piilo}
\affiliation{Department of Physics and Astronomy, University of Turku, FI-20014
 Turku, Finland}
\author{Kalle-Antti Suominen}
\affiliation{Department of Physics and Astronomy, University of Turku, FI-20014
 Turku, Finland}

\email{sabrina.maniscalco@utu.fi}

\date{\today}

\begin{abstract}
We derive and solve analytically the non-Markovian master equation for harmonic quantum Brownian motion proving that, for weak system-reservoir couplings and high temperatures, it can be recast in the form of the master equation for a harmonic oscillator interacting with a squeezed thermal bath. This equivalence guarantees preservation of positivity of the density operator during the time evolution and allows one to establish a connection between the dynamics of Schr\"odinger cat states in squeezed environments and environment-induced decoherence in quantum Brownian motion.
\end{abstract}

\maketitle

\section{Introduction}
The increasing ability in coherent control and manipulation of
the state of quantum systems has paved the way to experiments that are able
to monitor the transition from quantum superpositions, such as the
 Schr\"{o}dinger cat states, to classical statistical
mixtures \cite{engineerNIST,engNPRA}. The emergence of the
classical world from the quantum world, due to decoherence induced
by the environment, has been extensively investigated in the last
few decades both in connection to  the fundamental issues of quantum
theory and in relation to the emerging quantum technologies. The
fragile nature of quantum superpositions and entangled states, which are 
exploited, e.g., in quantum communication, quantum computation,
and quantum metrology, makes these potentially very powerful
techniques also very delicate \cite{BlattNat01}.

In order to understand the quantum to classical transition in terms of decoherence induced by the environment a paradigmatic model of the theory of open quantum systems, namely the quantum Brownian motion (QBM) model, has been extensively used in the literature \cite{Weiss99a,Breuer02a,Feynman63a,Caldeira83a,Haake85a,Grabert88a,Hu92a,Intravaia03a,EPJRWA,Maniscalco04b,Zurek,Eisert,Paz93,Maniscalco04a,JOPBManiscalco}. This model can be solved exactly using the path integral approach. This numerical method involves, however, complicated integrals and therefore simpler analytical approaches are desirable. In the weak-coupling limit, a secular master equation has been derived and investigated extensively in Refs. \cite{Intravaia03a,EPJRWA,Maniscalco04b,Maniscalco04a,JOPBManiscalco}. The secular approximation simplifies greatly the study of the dynamics and allows one to understand the origin of microscopic physical processes, such as the virtual exchanges of energy between the system and the environment, characterizing the non-Markovian dynamics. This approximation can be safely used when dealing with a class of observables that, in the weak coupling limit, do not depend on the non-secular terms, such as the mean energy of the system \cite{Intravaia03a,EPJRWA}. The study of the non-Markovian short time behavior of the density matrix of the quantum Brownian particle, however, requires in general the use of the complete QBM master equation. 

It is well known that the weak coupling limit of the QBM master equation is operatorially not in the Lindblad form  \cite{Breuer02a}. This fact complicates the analysis of the system dynamics and requires one to resort to numerical approaches. Moreover, the positivity of the density matrix during the time evolution can be violated, leading to unphysical results \cite{Haake96}. However, by adding some terms that are negligible in the weak coupling limit, the Markovian master equation can be rewritten in the Lindblad form (see, e.g., Ref. \cite{Breuer02a}). The main result of this paper is to establish a connection between the weak coupling master equation for QBM and the master equation for a harmonic oscillator interacting with a squeezed thermal reservoir \cite{Buzek92,Walls88,Kim93,Paris04,Paris03}. The latter one has been extensively studied in  the quantum optical context, and the conditions under which the positivity of the density matrix is preserved during the time evolution are known. Using the connection between these two master equations we obtain an analytic solution to the quantum Brownian motion dynamics for a weak coupling but in presence of a structured reservoir, i.e., without performing the Markovian approximation. Our analytic solution allows one to give a simple and physically clear description of environment-induced decoherence of a Schr\"odinger cat state for the QBM model, and provides results in agreement with Ref. \cite{Paz93}, in the Markovian limit. 

Very recently the quantum Brownian motion model has been used to study the entanglement between two harmonic oscillators in a common bath \cite{matteoeio,Horhammer,paz,xiang}.  In this context a very rich scenario showing the appearance of entanglement sudden death and revivals has been brought to light. We believe that our results may shed light on the microscopic origin of these phenomena.

The outline of the paper is the following. In Sec.~II we recall the master equation for quantum Brownian motion and its properties. In  Sec.~III we derive the weak coupling non-Markovian master equation and we establish the connection with the master equation describing a harmonic oscillator in a squeezed thermal bath. Section~IV is devoted to the solution of the master equation and to the study of the non-Markovian dynamics  for an initial Schr\"odinger cat state. Finally, Sec.~V contains the conclusions.

\section{The QBM Master equation}
We consider a harmonic oscillator linearly coupled to an engineered reservoir
modelled as an infinite chain of non-interacting
oscillators~\cite{Weiss99a,Breuer02a,Feynman63a,Caldeira83a,Haake85a,Grabert88a,Hu92a}.
The microscopic Hamiltonian of the total system, in units of
$\hbar$, has the form
\begin{eqnarray}
H &=& H_0+H_E + H_{\rm int}, \label{eqHtotal} \\
H_0 &=& \omega_0 \left( a^{\dag} a + \frac{1}{2} \right), \\
H_E &=& \sum_n \omega_n \left( b_n^{\dag} b_n + \frac{1}{2}
\right),\\
H_{\rm int} &=& g \sum_n \kappa_n \left( a^{\dag} + a \right)
\left( b_n^{\dag} + b_n \right), \label{eqHint}
\end{eqnarray}
where $H_0$, $H_E$, and $H_{\rm int}$ are the system, environment
and interaction Hamiltonians, respectively, $a$ ($a^{\dag}$) and
$b_n$ ($b^{\dag}_n$) are the annihilation (creation) operators of
the system and of the reservoir quantum oscillators, respectively,
$\omega_0$ is the frequency of the system oscillator, $\omega_n$
are the frequencies of the reservoir oscillators, $g$ is the
coupling constant, and the quantities $\kappa_n$ describe how
strongly the reservoir oscillators are coupled to the system. In
the continuum limit one introduces the spectral distribution $J(\omega)$ related to $\kappa_n$ via the equation $J(\omega) = \sum_n \kappa_n
\delta(\omega-\omega_n) / (2 m_n \omega_n)$, with $m_n$  as the mass for each environmental oscillator \cite{Breuer02a}.

Following the standard derivation of the master equation for the
reduced system (see, e.g., Ref.~\cite{Breuer02a}) one can demonstrate that the exact master equation, in the interaction picture, takes the form \cite{Hu92a,Intravaia03a,EPJRWA}
\begin{eqnarray}
 \frac{d \rho(t)}{dt} &=& 
 -   \Delta(t) [X,[X,\rho(t)]]  \nonumber \\
 &+& \Pi(t) [X,[P,\rho(t)]]+ \frac{i}{2} r(t) [X^2,\rho(t)] \nonumber \\
 &-& i \gamma(t) [X,\{P,\rho(t)\}], \label{Eq:MQbm}
\end{eqnarray}
where $\rho(t)$ is the reduced density matrix, $X = 
\left( a + a^{\dag}\right) /\sqrt{2 }$ and $P=i
\left( a^{\dag}- a\right)/\sqrt{2} $. 

The master equation given by Eq.~(\ref{Eq:MQbm}), being exact,
describes also the non-Markovian short time system-reservoir
correlations due to the finite correlation time of the reservoir.
In contrast to other non-Markovian dynamical systems, this master
equation is local in time, i.e., it does not contain memory
integrals. All the non-Markovian character of the system is
contained in the time-dependent coefficients appearing in the
master equation (for the analytic expression of the coefficients
see, e.g., Ref.~\cite{Maniscalco04b}). These coefficients depend
uniquely on the form of the reservoir spectral density. The
coefficient $r(t)$ describes a time-dependent frequency shift,
$\gamma(t)$ is the damping coefficient, $\Delta(t)$ and $\Pi(t)$
are the normal and the anomalous diffusion coefficients,
respectively \cite{Hu92a}.

The dynamics of the system has been extensively studied numerically using the path integral approach (see Refs. \cite{Grabert88a,Zurek} for a review). In particular, this model has been used to demonstrate the action of environment-induced decoherence for initial Schr\"{o}dinger cat states such as, e.g,
\begin{eqnarray}
\vert \Psi \rangle = \frac{1}{\sqrt{{\cal N}}} \left( \vert \alpha
\rangle + \vert - \alpha \rangle \right), \label{eq:inistate}
\end{eqnarray}
where $\vert \alpha \rangle$ is a
coherent state,
\begin{eqnarray}
{\cal N}^{-1}=2 \left[1+ \exp \left(-2 |\alpha|^2 \right) \right],
\end{eqnarray}
and we take  $\alpha \in \mathbb{R}$ for simplicity.
It has been proven that the decoherence induced by the environment acts in a much faster time scale than the thermalization process, in particular the decoherence time $\tau_d$ is inversely proportional to the separation between the two components of the superposition, i.e. $2|\alpha|^2$. For this reason, when the environmental spectrum has a structure, the reservoir correlation time, characterizing the duration of the non-Markovian dynamics, can become comparable to the decoherence time scale. In this case a non-Markovian description of environment-induced decoherence is important even in the weak coupling limit.

\section{Connection with the Master equation for a squeezed thermal bath}

For the sake of definiteness we focus on an Ohmic  reservoir described by a spectral distribution of the form \cite{Weiss99a}
\begin{eqnarray}
J(\omega)= \frac{2  \omega}{\pi} \
\frac{\omega_c^2}{\omega_c^2+\omega^2}, \label{Eq:SpecDen}
\end{eqnarray}
with $\omega_c$ as the cutoff frequency. The extension of our results to other forms of spectral densities is straightforward. In the limit of high temperatures and for sufficiently weak system-reservoir couplings, both the anomalous diffusion term $\Pi(t)$ and the frequency shift term $r(t)$ are negligible \cite{Paz93,Intravaia03a} and the QBM master equation~(\ref{Eq:MQbm}) can be recast in the form
\begin{eqnarray}
\frac{ d \rho(t)}{d t} &=& \frac{\Delta(t) \!+\! \gamma (t)}{2}
L(a^{\dag}) +\frac{\Delta(t) \!-\! \gamma (t)}{2} L[a] \nonumber \\
  &+& \frac{\Delta(t)}{2}e^{2 i \omega_0 t} D[ a^{\dag 2}] + \frac{\Delta(t)}{2} e^{-2 i \omega_0 t} D[  a^2],  \label{Eq:MEsqueez0}
\end{eqnarray}
where the superoperators $L[O]$ and $D[O]$ are given by
\begin{eqnarray}
L[O]&=&OO^{\dag} \rho + \rho OO^{\dag} -2 O^{\dag} \rho O, \nonumber \\
D[O] &=& O^2 \rho + \rho O^2 -2 O \rho O.
\end{eqnarray}
Equation (\ref{Eq:MEsqueez0}) is not in the Lindblad form, but has the same operatorial structure of the master equation for a harmonic oscillator in a thermal squeezed bath. Indeed, having in mind, e.g., Eq.~(6) of Ref.~\cite{Kim93} we can see that the two master equations coincide provided that we take $\Delta(t)+\gamma(t)=\gamma(N+1)$, $\Delta(t)-\gamma(t)\equiv \gamma  N$  and $\Delta(t) e^{2i \omega_0 t} \equiv - \gamma M$. The condition for positivity of the density operator, for a squeezed reservoir, reads as follows $|M|^2 \le N(N+1)$. It is straightforward to verify that Eq.~(\ref{Eq:MEsqueez0})  violates this condition since $|M|^2=\Delta (t)^2 \ge \Delta(t)^2 - \gamma(t)^2 = N(N+1)$.

In order to cure this problem we take a closer look at the second order expansion of the diffusion and dissipation coefficients appearing in Eq.~(\ref{Eq:MEsqueez0}) \cite{Hu92a,Maniscalco04b}
\begin{eqnarray}
\!\!\Delta(t)\!\!\!&=&\! \!g^2\!\! \!\!\int_0^t \!\!\!\!
\int_0^{\infty}\!\!\!\!\! \!\!d \omega d t_1 J(\!\omega\!)\!
\left[2N(\omega)+1\right] \!\cos(\omega t_1\!) \!\cos (\omega_0
t_1\!), \nonumber \\ \label{delta} \\
\gamma(t)&=& g^2 \int_0^t\!\! \int_0^{\infty}\!\!\! d \omega d
t_1  J(\omega) \sin(\omega t_1) \sin (\omega_0 t_1), \label{gamma}
\end{eqnarray}
where  $N(\omega) =
(e^{ \omega/k_B T}-1)^{-1}$ is the average number of reservoir
thermal photons, $k_B$ is the Boltzmann  constant, and $T$ is the
reservoir temperature. We note that, for high $T$, i.e., $N(\omega)
\gg 1$, we have $\Delta(t)\gg \gamma(t)$.

Having this in mind we can modify the weak coupling master equation  (\ref{Eq:MEsqueez0}) by adding some terms, proportional to $\gamma(t)$, which are negligible with respect to the terms proportional to $\Delta(t)$ in the high $T$ and weak coupling limits. The resulting master equation takes the form
\begin{eqnarray}
\frac{ d \rho(t)}{d t} &=& \frac{\Delta(t) \!+\! \gamma (t)}{2}
L(a^{\dag}) +\frac{\Delta(t) \!-\! \gamma (t)}{2} L[a] \nonumber \\
  &+& \frac{\Delta(t)-\gamma(t)}{2} e^{2 i \omega_0 t} D[ a^{\dag 2}] \nonumber \\
  &+& \frac{\Delta(t)-\gamma(t)}{2} e^{-2 i \omega_0 t} D[ a^2]. \label{Eq:MEsqueez}
\end{eqnarray}
In this case $\Delta(t)+\gamma(t)=\gamma(N+1)$, $\Delta(t)-\gamma(t)\equiv \gamma  N$  and $[\Delta(t)-\gamma(t)] e^{2i \omega_0 t} \equiv - \gamma M$, and the condition for positivity of the density operator $|M|^2 \le N(N+1)$ is satisfied as an equality. This corresponds to the case of a maximally squeezed reservoir ($|M|=N$).

The analytical form of the coefficients $\Delta(t)$ and $\gamma(t)$ can be easily calculated explicitly for an Ohmic environment, in the high $T$ limit, inserting Eq.~(\ref{Eq:SpecDen}) into Eqs. ~(\ref{delta})-(\ref{gamma}),
\begin{eqnarray}
\Delta(t) &=& 2 g^2 k_B T \frac{r^2}{1+r^2} \left\{ 1
- e^{-\omega_c t} \left[ \cos (\omega_0 t)\right.\right. \nonumber \\
&-&  (1/r) \left. \left. \sin (\omega_0 t )\right] \right\},
\label{deltaHT} \\
\gamma (t)\! \!&=&\!\! \frac{g^2 \omega_0 r^2}{1+r^2} \Big[1
\!-\! e^{- \omega_c t} \cos(\omega_0 t) \! - r e^{- \omega_c t}
\sin( \omega_0 t )  \Big], \nonumber \\ \label{gammasecord}
\end{eqnarray}
with $r=\omega_c/\omega_0$.

From these equations we see that $\Delta(t)$ and $\gamma(t)$ start from an initial zero value and quickly approach their constant Markovian value. Indeed for $t \gg \tau_R=1/\omega_c$, with $\tau_R$ as the reservoir correlation time, 
the time-dependent coefficients $\Delta(t)+\gamma(t)$ and
$\Delta(t)- \gamma(t)$ become
\begin{eqnarray}
\Delta(t)+\gamma(t) &\simeq& \Gamma [N(\omega_0)+1] \equiv \gamma_{1}^M  , \label{eq:marg1}\\
\Delta(t) - \gamma (t) &\simeq&  \Gamma N(\omega_0) \equiv \gamma_{-1}^M, \label{eq:margm1}
\end{eqnarray}
respectively, with $N(\omega_0) \simeq k_B T /\omega_0$ and
\begin{eqnarray}
\Gamma = 2 g^2 \frac{r^2}{r^2+1} \omega_0. \label{eq:Gamma}
\end{eqnarray}

In the next section we will present the solution of this master equation~(\ref{Eq:MEsqueez}) in terms of the Wigner function and discuss its properties for an initial state of the form of Eq.~(\ref{eq:inistate}). Moreover, we will check the validity of the approximations under which this master equation holds by comparing the solution we have derived with the solution of the exact Hu-Paz-Zhang master equation~(\ref{Eq:MQbm}).

By looking at Eq.~(\ref{Eq:MEsqueez}) one realizes immediately that, for times $\omega_0 t \gg 1$ the last two terms average out to zero and the master equation reduces to the secular master equation used, e.g., in Refs.~\cite{Maniscalco04b,Maniscalco04a,Piilo07a,Maniscalco06a}
\begin{eqnarray}
&&\frac{ d \rho(t)}{d t}= \frac{\Delta(t) \!+\! \gamma (t)}{2}
\left[2 a \rho(t) a^{\dag}- a^{\dag} a \rho(t)  - \rho(t) a^{\dag}
a \right]
\nonumber \\
&& +\frac{\Delta(t) \!-\! \gamma (t)}{2} \left[2 a^{\dag} \rho(t)
a - a a^{\dag} \rho(t) - \rho(t) a a^{\dag}
 \right]. \nonumber \\
 \label{Eq:MQsecapprox}
\end{eqnarray}
It is known that in the weak coupling limit there exists a class of observables, e.g. $n=a^{\dag}a$, whose dynamics is not affected by the counter-rotating terms, i.e. by those terms neglected in the secular approximation \cite{Grabert88a,Intravaia03a}. In general, however, the counter-rotating terms do contribute to the dynamics of the reduced density operator of the system, in particular in the short non-Markovian time scale we are interested in. 

In the following we will focus on two specific physical regimes characterized by opposite values of the parameter $r$, namely the case in which $r = \omega_c/\omega_0 \gg1$ and the  case $r \ll 1$. We will refer to these cases as the resonant and the off-resonant case, respectively, since for $r \gg 1$ the frequency $\omega_0$ of the oscillator overlaps with the spectrum of the reservoir while, for $r\ll1$, $\omega_0$ is off-resonant with the spectrum $J(\omega)$. From previous studies on the dynamics of the heating function $\langle n (t) \rangle$ we know that the system time evolution differs notably in these two regimes \cite{Maniscalco04b}. As we will see in Sec. IV, also the dynamics of an initial Schr\"odinger cat state crucially depends on the value of the resonance parameter $r$.

It is worth emphasizing that, in the resonant case $r \gg 1$,
for $\omega_c t \le 1$, we cannot use the secular master equation  since   $\omega_0 t \ll \omega_c t \le 1$. Therefore we should use Eq.~(\ref{Eq:MEsqueez}). On the contrary, in the off-resonant case $r \ll 1$, we can focus on the dynamics for times $1/\omega_0 \ll t \le 1/\omega_c$, since this is consistent with the assumption $r=\omega_c/\omega_0 \ll 1$, and use the simpler secular master equation~(\ref{Eq:MQsecapprox}).

\section{The dynamics}
We now look at the short-time  dynamics of  a Schr\"{o}dinger cat state of the form given in Eq.~(\ref{eq:inistate}).
This state is also known as even coherent state due to the fact
that only the even components of the number probability
distribution are nonzero. The oscillations in the number state
probability are a strong sign of the nonclassicality of this
state. This and other nonclassical properties of the even coherent
state, such as the negativity of the corresponding Wigner
function, have been extensively studied in the literature (see,
e.g., Ref.~\cite{Buzek92} and references therein). In Fig.~1 we show the number probability distribution and the Wigner function for
the even coherent state. This state has been realized in the trapped ion context and the transition from a quantum superposition to a classical statistical mixture has been observed experimentally~\cite{engineerNIST}.

The decoherence and dissipation due to the interaction with both thermal
reservoirs and squeezed reservoirs has been studied, in the Markovian limit, in Refs.~\cite{Buzek92} and \cite{Walls88,Kim93,Paris04}, respectively. 
On the other hand, the exact dynamics of the Hu-Paz-Zhang master equation~(\ref{Eq:MQbm}) has been numerically investigated in Refs.~\cite{Paz93,Zurek}. In the following we will look for a simple analytic solution valid in the short-time non-Markovian regime in the two limits $r\ll1$ and $r\gg1$. We will verify the validity of the approximations made to derive such a solution comparing it with the Hu-Paz-Zhang result~\cite{Paz93,Zurek}.

\begin{figure}
\centering
\includegraphics[scale=0.4]{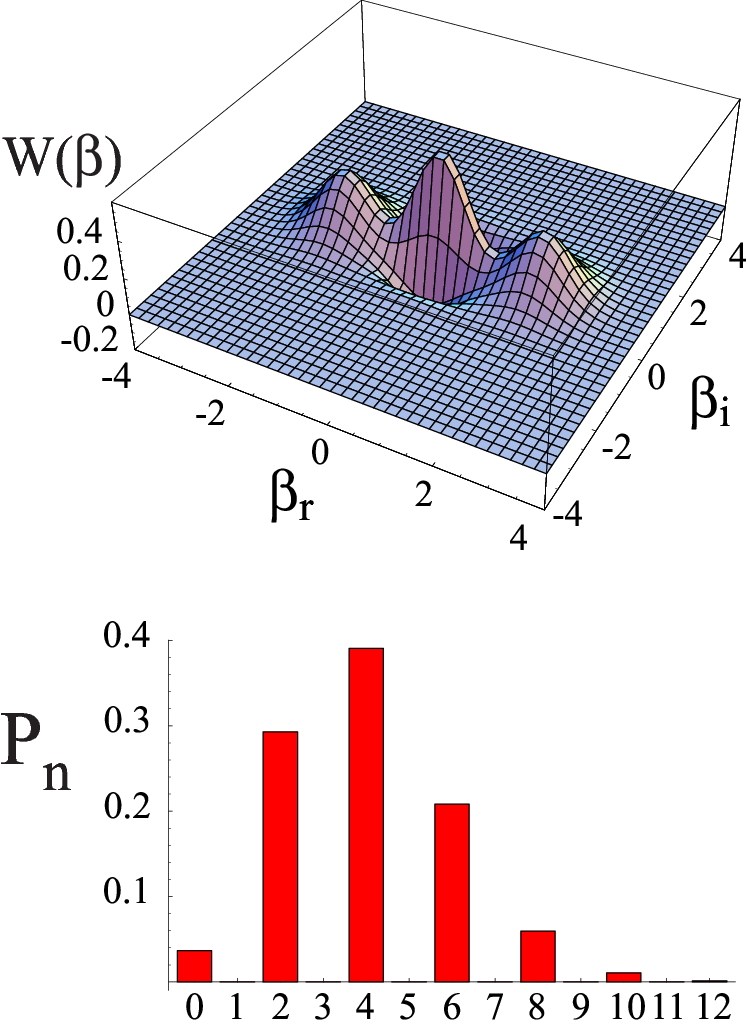}
\caption{\label{Fig1} (Color on line) Wigner function $W(\beta)$
and number probability distribution $P_n$ for the state given by
Eq.~(\ref{eq:inistate}) with $\alpha = 2$.}
\end{figure}

In order to describe the transition from the initial even coherent state to the corresponding classical statistical mixture induced by the interaction with the environment it is convenient to look at the dynamics of the Wigner funtion.  In order to investigate the non-Markovian dynamics we need to find out the time evolution of the Wigner function for times $t \le 1/\omega_c$. 

\subsubsection{The off-resonant case $r\ll1$}

In the off-resonant case $r \ll 1$ the solution of Eq.~(\ref{Eq:MQsecapprox})  in terms of the quantum characteristic function (QCF)  $\chi(\xi)$ is presented in Ref.~\cite{Intravaia03a}. From this solution, and remembering that the Wigner function is the
Fourier transform of the QCF,
\begin{eqnarray}
W(\beta)= \frac{1}{\pi^2} \int_{-\infty}^{\infty} d^2 \xi
\chi(\xi) \exp \left( \beta \xi^* - \beta^* \xi  \right),
\end{eqnarray}
one obtains straightforwardly the following expression for the dynamics of an initial even coherent state.

\begin{eqnarray}
W(\beta ,t) =  W^{(+ \alpha)}(\beta ,t) +  W^{(-\alpha)}(\beta ,t) +
W_I(\beta ,t), 
\end{eqnarray}
with
\begin{eqnarray}
\label{eq:Wigner1a} W^{\pm \alpha}(\beta , t)&=& \frac{ {\cal N}}{\pi \left[N(t)+1/2
\right]} \exp \left( - \frac{ \beta_i^2}{N(t)+1/2}\right) \nonumber \\
&\times& \exp \left[ - \frac{\left( \beta_r \mp e^{-\Gamma(t)/2}
\alpha \right)^2}{N(t)+1/2} \right], \label{Wmix} \\
\label{eq:Wigner1b} W_I(\beta ,t) &=& \frac{ {2 \cal N}}{\pi \left[ N(t)+1/2 \right]}
\exp \left( - \frac{ |\beta|^2}{
N(t)+1/2}\right) \nonumber \\
&\times& \exp \left[ -2 \alpha^2 \left( 1-\frac{e^{-\Gamma(t)} }{
2 N(t)+1}\right)  \right] \nonumber \\
&\times& \cos \left[ \frac{2 e^{- \Gamma(t)/2}}{N(t)+1/2} \alpha
\beta_i \right], \label{Eq:wignersaI}
\end{eqnarray}
with 
\begin{eqnarray}
N(t)&=&\int_0^t dt' \Delta(t'), \label{Eq:N} \\
\Gamma(t) &=& 2 \int_0^t dt' \gamma(t'), \label{Eq:Gammat}
\end{eqnarray}
where the coefficients $\Delta(t)$ and $\gamma(t)$ are given by Eqs.~(\ref{deltaHT})-(\ref{gammasecord}). As known from the Markovian theory, the interaction with the environment causes the disappearance of the interference peak and therefore the transition from quantum superposition to classical statistical mixture. A useful quantity to monitor this transition is the fringe visibility function 
\begin{eqnarray}
 F(\alpha,t) &\equiv& \exp (-A_{int}) \label{Eq:fringe} \\ \nonumber
&=& \frac{1}{2} \frac{\left. W_I( \beta,t) \right|_{\rm peak}}{\left[ \left. W^{(+\alpha)} (\beta,t) \right|_{\rm peak}   \left. W^{(-\alpha)} (\beta,t) \right|_{\rm peak} \right]^{1/2}},
\end{eqnarray}
where we indicate with $\left. W_I( \beta,t) \right|_{\rm peak}$ and $\left. W^{(\pm \alpha)} (\beta,t) \right|_{\rm peak} $ the value of the Wigner function at $\beta=(0,0)$ and $\beta=(\pm \alpha,0)$, respectively.
Inserting Eqs.~(\ref{Wmix})-(\ref{Eq:wignersaI}) into Eq.~(\ref{Eq:fringe}) we obtain, for the $r\ll 1$ case, 
\begin{eqnarray}
F(\alpha,t) =  \exp \left[ -2 \alpha^2 \left( 1-\frac{e^{-\Gamma(t)} }{
2 N(t)+1}\right)  \right]. \label{Eq:fringersmall}
\end{eqnarray} 

This equation tells us that, as for the Markovian theory, the interference term disappears  faster and faster, the larger is the separation between the two components of the superposition, measured by $2 \alpha$, and the higher is the temperature of the environment [see Eqs.~(\ref{deltaHT})-(\ref{Eq:N})]. In Fig. 2 we plot the time evolution of the fringe visibility factor comparing the off-resonant (dashed line) and the on-resonant (solid line) cases.

\subsubsection{The resonant case $r \gg1$}

We now consider the dynamics in the more complicated resonant case. We begin by noting that, for $\omega_c t \le 1$, we can approximate $e^{2i \omega_0 t} \simeq 1$ since $r = \omega_c / \omega_0 \gg 1$. As a consequence we have $M \in \mathbb{R}$ and $M=-N$. 

The solution of the master equation~(\ref{Eq:MEsqueez}) can be obtained following the same steps as in solving the master equation for a harmonic oscillator in a thermal squeezed bath, in the Markovian case. We write down the Fokker-Planck equation corresponding to the master equation~(\ref{Eq:MEsqueez}) for the Wigner function. Since the initial state is a linear combination of Gaussian terms, the form of the master equation ensures that each Gaussian term evolves independently. Therefore the evolved state will also be a linear combination of Gaussian terms~\cite{Paris03}.  Having this in mind it is straightforward to derive the Wigner function dynamics as follows 
\begin{eqnarray}
\label{eq:Wigner2a} W^{\pm \alpha}(\beta , t)&=& \frac{ {\cal N}}{\pi\left\{ N(t)+1/4 \right\}^{1/2}} \nonumber \\
&\times& \exp \left( - \frac{ \beta_i^2}{2N(t)+1/2}\right) \nonumber \\
&\times& \exp \left[ - \frac{\left( \beta_r \mp e^{-\Gamma(t)/2}
\alpha \right)^2}{1/2} \right], \label{Wmixsque} \\
\label{eq:Wigner2b} W_I(\beta ,t) &=& \frac{ {2 \cal N}}{\pi \left\{ N(t)+1/4\right\}^{1/2}} \nonumber \\
&\times&
\exp \left( - \frac{ \beta_i^2}{
2N(t)+1/2} - \frac{ \beta_r^2}{
1/2}\right) \nonumber \\
&\times& \exp \left[ -2 \alpha^2 \left( 1-\frac{e^{-\Gamma(t)} }{
4 N(t)+1}\right)  \right] \nonumber \\
&\times& \cos \left[ \frac{2 e^{- \Gamma(t)/2}}{2N(t)+1/2} \alpha
\beta_i \right], \label{Eq:wignersque}
\end{eqnarray}
with $N(t)$ and $\Gamma(t)$ given by Eqs.~(\ref{Eq:N})-(\ref{Eq:Gammat}). Comparing the dynamics to the case $r\ll 1$, where the secular approximation holds, we see the following differences. First of all, the variances of the two Gaussians $W^{\pm \alpha}(\beta,t)$, correspondent to the two components of the superposition, do not follow the same dynamics anymore. This asymmetry in the time evolution of the variances is typical of the behavior of harmonic oscillators in squeezed environments. But more interesting for the study of the quantum-classical transition is the behavior of the fringe visibility 
\begin{eqnarray}
F(\alpha,t) =  \exp \left[ -2 \alpha^2 \left( 1-\frac{e^{-\Gamma(t)} }{
4 N(t)+1}\right)  \right]. \label{Eq:fringerbig}
\end{eqnarray} 
Comparing this equation with Eq.~(\ref{Eq:fringersmall}) obtained for $r \ll1$ we notice that, with regard to the decoherence process, in the resonant case $r \gg 1$ it is as if the system would interact with a thermal reservoir with an effective temperature that is the double of the real temperature. This is simply due to the non-negligible role played by the counter-rotating terms present in the microscopic interaction Hamiltonian of Eq.~(\ref{eqHint}). In Fig. 2 we plot the time evolution of the fringe visibility factor for an initial Schr\"odinger cat state with $\alpha=4$, in the cases $r=10$ (solid line) and $r=0.1$ (dashed line).

\begin{figure}
\centering
\includegraphics[scale=0.6]{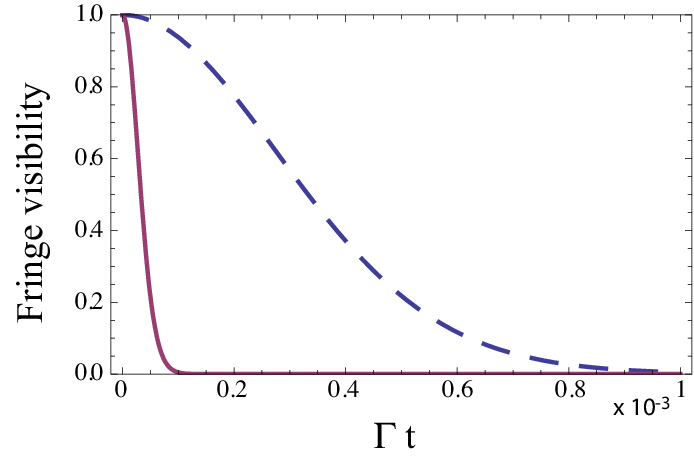}
\caption{\label{Fig2} (Color on line) Time evolution of the fringe visibility factor for the state given by
Eq.~(\ref{eq:inistate}) with $\alpha = 4$, $k_B T/ \omega_0 = 100$,  $r=10$ (solid line) and $r=0.1$ (dashed line).}
\end{figure}

To conclude this section we compare the dynamics of the fringe visibility factor obtained without performing the secular approximation, i.e.  by solving Eq.~(\ref{Eq:MEsqueez}), with the result presented in Ref.~\cite{Paz93} for the exact Hu-Paz-Zhang master equation~(\ref{Eq:MQbm}). Combining Eq.~(\ref{Eq:fringe}) and Eq.~(\ref{Eq:fringerbig}), and using Eq.~(\ref{Eq:N}) we obtain
\begin{eqnarray}
A_{int}= 2 \alpha^2 \left( \frac{4 \int_0^t dt' \Delta(t')}{4 \int_0^t dt' \Delta(t') + 1} \right), \label{Eq:Aint}
\end{eqnarray}
where we have put $e^{-\Gamma(t)} \simeq 1$ since we are far from thermalization, i.e., $t \ll t_{th}$. We recall that after a time $t \simeq \tau_R$ the diffusion coefficient attains its constant Markovian value $\Delta_M$ and therefore $ \int_0^t dt' \Delta(t') \simeq \Delta_M t$. Having this in mind we can compare Eq.~(\ref{Eq:Aint}) with the high temperature and low damping approximation of $A_{int}$ given by Eq.~(42) of  Ref.~\cite{Paz93}. One sees immediately that, mutatis mutandi, our Eq.~(\ref{Eq:Aint}) coincides with Eq.~(42) of  Ref.~\cite{Paz93} for $t \gg \tau_R$ and also it provides a straightforward generalization of this equation for the case $t \le \tau_R$.

Another important aspect, discussed in  Ref.~\cite{Paz93} and predicted by the exact  Hu-Paz-Zhang model, is the different decoherent dynamics of superpositions of wave packets separated in positions, such as those we considered in our example, and of superposition of wave packets separated in momentum. The latter ones experience a much slower decoherence compared to the former ones. In Ref.~\cite{Paz93} this fact was attributed to the microscopic interaction Hamiltonian coupling the position operator $X$ of the system oscillator to the bath. This prediction is confirmed by our model. For times $t \ll t_{\rm therm}$, with $t_{\rm therm}$ the thermalization time, indeed, we can neglect the small difference between the rates $\Delta(t)+\gamma(t)$ and $\Delta(t)-\gamma(t)$, corresponding to the upward and downward transition, i.e., to the absorption and emission of a quantum of energy by the system oscillator, respectively. In this case the master eqution becomes

\begin{eqnarray}
\frac{ d \rho(t)}{d t} &=& \frac{\Delta(t) }{2}
L(a^{\dag}) +\frac{\Delta(t) }{2} L[a] \nonumber \\
  &+& \frac{\Delta(t)}{2}e^{2 i \omega_0 t} D[ a^{\dag 2}] + \frac{\Delta(t)}{2} e^{-2 i \omega_0 t} D[  a^2] \nonumber \\
  &=& \Delta(t) [X,[X,\rho]],  \label{Eq:MEsqueezap}
\end{eqnarray}
describing a continuous measurement of the position of the system harmonic oscillator.

\section{Summary and Conclusions}

In this paper we have studied the non-Markovian short time dynamics of the harmonic QBM model.
In the high temperature case we have obtained a weak coupling master equation preserving positivity of the density matrix and able to describe situations where the spectrum of the environment is not flat. This master equation has the same form as the master equation describing  a harmonic oscillator interacting with a squeezed thermal reservoir and preserves positivity of the density matrix during the time evolution, thus solving the problem discussed, e.g., in Ref. \cite{Haake96}.

Using this analogy, we have obtained a solution in terms of the Wigner function and investigated the dynamics in two opposite physical regimes characterized by the parameters $r \gg 1$ (resonant case) and $r \ll 1$ (off resonant case). A comparison between the two solutions reveals that the counter-rotating terms present in the microscopic Hamiltonian model have a non-negligible effect in the resonant case. On the contrary, in the off-resonant case they can be neglected.

The availability of simple analytic expressions describing the short time dynamics of the system oscillator has allowed us to shed light on the non-Markovian dynamics of environment induced decoherence and, in particular, on its sensitivity to the form of the initial superposition.

Because of the fragility of such states it is interesting to see whether it is possible to modify the decoherence induced by the environment, e.g., by means of the quantum Zeno effect \cite{Maniscalco06a}. We plan to investigate this issue in the future.

\section*{Acknowledgements}
The authors acknowledge financial support from the Academy of
Finland (projects 108699, 115982, 115682), the Magnus Ehrnrooth
Foundation and the V\"{a}is\"{a}l\"{a} Foundation.

\end{document}